\documentclass[twocolumn,showpacs,amsmath,amssymb]{revtex4}
\input{epsf}

\usepackage{graphicx}
\usepackage{array}
\usepackage{bigstrut}
\usepackage{longtable}
\usepackage{rotating,booktabs}
\usepackage{booktabs,threeparttable}
\usepackage{bm}
\usepackage{lipsum}

\begin{document}

\title{ Relativistic full-configuration-interaction calculations of magic wavelengths for the $2\,^3S_1\rightarrow2\,^1S_0$ transition of helium isotopes }

\author{Fang-Fei Wu$^{1,2}$, San-Jiang Yang$^{1,3}$, Yong-Hui Zhang$^{1,*}$~\footnotetext{*Email Address: yhzhang@wipm.ac.cn}, Jun-Yi Zhang$^{1}$, Hao-Xue Qiao$^3$, Ting-Yun Shi$^{1,4}$, and Li-Yan Tang$^{1,\dag}$~\footnotetext{\dag Email Address: lytang@wipm.ac.cn}}

\affiliation {$^1$State Key Laboratory of Magnetic Resonance and
Atomic and Molecular Physics, Wuhan Institute of Physics and
Mathematics, Chinese Academy of Sciences, Wuhan 430071, People's Republic of China}

\affiliation {$^2$University of Chinese Academy of Sciences, Beijing 100049, People's Republic of China}

\affiliation {$^3$Department of Physics, Wuhan University, Wuhan 430072, People's Republic of China}

\affiliation {$^4$ Center for Cold Atom Physics, Chinese Academy of Sciences,
Wuhan 430071, People¡¯s Republic of China}

\date{\today}

\begin{abstract}
A large-scale full-configuration-interaction calculation based on Dirac-Coulomb-Breit (DCB) Hamiltonian is performed for the $2\,^1S_0$ and $2\,^3S_1$ states of helium. The operators of the normal and specific mass shifts are directly included in the DCB framework to take the finite nuclear mass correction into account. High-accuracy energies and matrix elements involved n (the main quantum number) up to 13 are obtained from one diagonalization of Hamiltonian. The dynamic dipole polarizabilities are calculated by using the sum rule of intermediate states. And a series of magic wavelengths with QED and hyperfine effects included for the $2\,^3S_1\rightarrow2\,^1S_0$ transition of helium are identified. In addition, the high-order Ac Stark shift determined by the dynamic hyperpolarizabilities at the magic wavelengths are also evaluated. Since the most promising magic wavelength for application in experiment is 319.8 nm, the high-accuracy magic wavelength of 319.815 3(6) nm of $^4$He is in good agreement with recent measurement value of 319.815 92(15) nm [Nature Physics (2018)/arXiv:1804.06693], and present magic wavelength of 319.830 2(7) nm for $^3$He would provide theoretical support for experimental designing an optical dipole trap to precisely determine the nuclear charge radius of helium in future.
\end{abstract}

\pacs{31.15.ap, 31.15.ac, 32.10.Dk} \maketitle

The long-term outstanding proton radius puzzle causes great interest in recent years~\cite{pohl10a,antognini13a,mohr16a}. So far there has not been a satisfying explanation for the discrepancy of 5.6$\sigma$ in the proton size derived from muonic hydrogen Lamb shift measurements~\cite{pohl10a,antognini13a} and the accepted CODATA value~\cite{mohr16a}. Research in this field has expanded to measurements of the 2$S$-4$P$ transition energy in hydrogen~\cite{beyer17a}, transition energies between circular Rydberg states in heavy-H-like ions~\cite{tan11a}, and the 1$S$-2$S$ transition energy in muonic helium ions~\cite{antognini11a}. In order to help solve the proton size puzzle, the measurement of high-precision spectroscopy in helium isotopes has become an additional contribution to this field~\cite{rooij11a,leeuwen06a,shiner95a,morton06a,pastor04a,pastor12a}. However, the nuclear charge radius difference determined from the $2\,^3S\rightarrow2\,^1S$ and $2\,^3S\rightarrow2\,^3P$ transitions of helium disagrees by 4$\sigma$~\cite{rooij11a,leeuwen06a,shiner95a, morton06a,pastor04a,pastor12a}. Even combined with the recent theoretical investigations~\cite{pachucki15a,patkos16a,patkos17a}, where the higher-order recoil corrections are taken into account, the 4$\sigma$ discrepancy does still exist and remains unexplained by any missed corrections in existing theoretical predictions. So this discrepancy calls for the verification of the experimental transition frequencies by independent measurements.

For the $2\,^3S\rightarrow2\,^3P$ transition frequency of helium, recently, the frequency measurement of $^4$He is achieved to $5.1\times 10^{-12}$~\cite{zheng17a}, which is more accurate than the early results of Refs.~\cite{pastor04a,pastor06a,pastor12a}. But it's interesting that applying the $2\,^3S\rightarrow2\,^3P$ transition frequency of $\,^3$He of Ref.~\cite{pastor12a} into Ref.~\cite{zheng17a}, the resulting nuclear charge radii difference agrees well with the value derived from the $2\,^3S\rightarrow2\,^1S$ transition~\cite{rooij11a} but differs with the measurement of $2\,^3S\rightarrow2\,^3P$ transition~\cite{shiner95a}. This deviation indicates that further independent measurement of the $2\,^3S\rightarrow2\,^3P$ transition for $^3$He is urgently needed.

For the $2\,^3S\rightarrow2\,^1S$ transition frequency of helium, one of the main systematic uncertainty of previous measurement~\cite{rooij11a} comes from Ac Stark shift. Implementation of a magic wavelength trap can solve this problem in many high-precision measurements~\cite{kim17a,campbell17a}. Recently, Notermans~\textit{et al.} obtain the magic wavelengths of He($2\,^3S\rightarrow2\,^1S$) with use of available energies and Einstein A coefficients~\cite{notermans14b}. The accuracy of their values are limited by extrapolated contributions from continuums.

Since the dynamic dipole polarizability at the 319.8 nm magic wavelength is large enough to provide sufficient trap depth at reasonable laser powers while the scattering lifetime is accepted, the 319.8 nm magic wavelength is proposed to design a optical dipole trap (ODT) to eliminate the Ac Stark shift~\cite{notermans14b}. In order to determine the nuclear charge radius difference with a precision comparable to the muonic helium ion, Vassen~\textit{et al.} aim to measure the $2\,^3S\rightarrow2\,^1S$ transition with sub-kHz precision. At this level of precision, the \textit{ab-initio} calculation for the magic wavelengths of helium isotopes are required.

In this paper, we improve the previous relativistic configuration interaction (RCI) method~\cite{zhang16a} by adding the mass shift (MS) operators directly into the Dirac-Coulomb-Breit (DCB) Hamiltonian. Then we perform a large-scale full-configuration-interaction calculation of the dynamic dipole polarizabilities for the $2\,^3S$ and $2\,^1S$ states of helium. QED corrections to the dynamic dipole polarizabilities are approximate by perturbation calculations. A series of magic wavelengths for the $2\,^3S_1\rightarrow2\,^1S_0$ transition are accurately identified according to the dynamic polarizabilities. In addition, we also carry out a non-relativistic calculations of dynamic polarizabilities and hyperpolarizabilities of helium by using the newly developed Hylleraas-B-spline method~\cite{yang17a}. Present magic wavelengths from two different theoretical methods are in good agreement. Specially, the accurate magic wavelengths of 319.815 3(6) nm and 319.830 2(7) nm are recommended for the $2\,^3S_1(M_J=\pm 1)\rightarrow2\,^1S_0$ transition of $^4$He and $^3$He, respectively.

The DCB Hamiltonian with mass shift operator included for the two-electron atomic system is written as
\begin{widetext}
\begin{eqnarray}
H=\sum\limits_{i=1}^{2}\left[c\bm{\alpha}_i\cdot\bm{p}_i+\beta m_ec^2-\dfrac{Z}{r_i}\right]+\dfrac{1}{r_{12}}
-\dfrac{1}{2r_{12}}\big[\bm{\alpha}_1\cdot\bm{\alpha}_2+\left(\bm{\alpha}_1\cdot\hat{\bm{r}}_{12}\right)
                                                             \left(\bm{\alpha}_2\cdot\hat{\bm{r}}_{12}\right)\big]+H_{MS} \,
\label{e1}
\end{eqnarray}
\end{widetext}
where $c=137.035999074$ is the speed of light~\cite{mohr12a}, $Z$ is the nuclear charge, $\beta$ is the $4\times4$ Dirac matrix, $m_e=1$ is the electron mass, $\bm{\alpha}_i$ and $\bm{p}_i$ are respectively the Dirac matrix and the momentum operator for the $i$-th electron, $\hat{\bm{r}}_{12}$ is the unit vector of the electron-electron distance $\bm{r}_{12}$, and the MS operator $H_{MS}$ includes the leading term of normal and specific mass shift (NMS, SMS) operators,
\begin{eqnarray}
H_{MS}=H_{NMS}+H_{SMS}=\sum\limits_{i=1}^{2}\dfrac{\bm{p}_i^2}{2m_0}+\dfrac{\bm{p}_1\cdot\bm{p}_2}{m_0}
\,,\label{e2}
\end{eqnarray}
where $m_0=7294.2995361\, m_e$ and $m_0=5495.8852754\, m_e$~\cite{mohr12a} are the nuclear mass for $^4$He and $^3$He, respectively. The wave function $\psi(JM_J)$ for a state with angular momentum $\left(J, M_J\right)$ is expanded as a linear combination of the configuration-state wave functions $\phi_{ij}(JM_J)$, which are constructed by the single-electron wave functions~\cite{zhang16a}. Using the Notre Dame basis sets~\cite{johnson88a, beloy08a} of B-spline functions, the single-electron wave functions are obtained by solving the single-electron Dirac equation.

The non-relativistic Hamiltonian for the infinite nuclear mass of helium is solved with Hylleraas-B-spline basis set~\cite{yang17a},
\begin{eqnarray}
\phi_{ij\ell_1\ell_2}\left(LM_L\right)&=&B_i^k(r_1)B_j^k(r_2)r_{12}^cY_{\ell_1\ell_2}^{LM_L}\left(\hat{r}_1,\hat{r}_2\right) \nonumber \\
&&\pm \, exchange \,,\label{e3}
\end{eqnarray}
where $L$ and $M_L$ are the total orbital and magnetic quantum numbers, respectively, $Y_{\ell_1\ell_2}^{LM_L}\left(\hat{r}_1,\hat{r}_2\right)$ is the coupled spherical harmornic function, and $c = 0\,, 1$.

The dynamic dipole polarizability of the magnetic sublevel $|N_gJ_gM_{J_g}\rangle$ under linear polarized light with laser frequency $\omega$ is
\begin{eqnarray}
\alpha_1(\omega)=\alpha_1^S(\omega)+\dfrac{3M_{J_g}^2-J_g(J_g+1)}{J_g(2J_g-1)}\alpha_1^T(\omega)
\,,\label{e4}
\end{eqnarray}
where $\alpha_1^S(\omega)$ and $\alpha_1^T(\omega)$ are the scalar and tensor dipole polarizabilities, respectively, which can be expressed as the summation over all intermediate states,
\begin{eqnarray}
\alpha_1^S(\omega)=\sum_{n\neq g}\dfrac{f_{gn}^{(1)}}{(\Delta E_{gn})^2-\omega^2}
\,,\label{e5}
\end{eqnarray}
\begin{widetext}
\begin{eqnarray}
\alpha_1^T(\omega)=\sum_{n\neq g}
(-1)^{J_g+J_n}\sqrt{\dfrac{30(2J_g+1)J_g(2J_g-1)}{(2J_g+3)(J_g+1)}}
\left\{
\begin{array}{ccc}
1   &1   &2\\
J_g &J_g &J_n\\
\end{array}
\right\} \dfrac{f_{gn}^{(1)}}{(\Delta E_{gn})^2-\omega^2}
\,,\label{e6}
\end{eqnarray}
\end{widetext}
with $f_{gn}^{(1)}$ is the dipole oscillator strength,
\begin{eqnarray}
f_{gn}^{(1)}=\dfrac{2|\langle N_gJ_g\|T_{1}
                                \|N_nJ_n\rangle|^2\Delta E_{gn}}{3(2J_g+1)}
\,,\label{e7}
\end{eqnarray}
where $\triangle E_{gn}=E_n-E_g$ is transition energy between the initial state $|N_gJ_g\rangle$ and the intermediate state $|N_nJ_n\rangle$,  and $T_1$ is the dipole transition operator.
The nonrelativistic polarizabilities are obtained by replacing the quantum number $J$ with $L$ in Eqs.(\ref{e4})- (\ref{e7}).

The QED corrections to dynamic dipole polarizabilities are calculated by the perturbation theory~\cite{pachucki00a} using energies and wavefunctions obtained from non-relativistic configuration interaction (NRCI) method~\cite{zhang15},
\begin{widetext}
\begin{eqnarray}
\delta \alpha_1^{QED}(\omega)&=&2\Bigg[
\sum\limits_n\dfrac{\langle g|T_1|n\rangle\langle n|T_1|g\rangle\langle g|\delta H_{QED}|g\rangle[(E_n-E_g)^2+\omega^2]}{[(E_n-E_g)^2-\omega^2]^2}-2\sum\limits_{nm}\dfrac{\langle g|T_1|n\rangle\langle n|T_1|m\rangle\langle m|\delta H_{QED}|g\rangle(E_n-E_g)}{[(E_n-E_g)^2-\omega^2](E_m-E_g)}  \nonumber \\
&-&\sum\limits_{nm}\dfrac{\langle g|T_1|n\rangle\langle n|\delta H_{QED}|m\rangle\langle m|T_1|g\rangle[(E_n-E_g)(E_m-E_g)+\omega^2]}{[(E_n-E_g)^2-\omega^2][(E_m-E_g)^2-\omega^2]} \Bigg]
\,\label{qed1}
\end{eqnarray}
\end{widetext}
where $|g\rangle$ is the initial state, $|n\rangle$ and $|m\rangle$ represent intermediate states, and $\delta H_{QED}$ is the QED operator. The expansion of $\delta H_{QED}$~\cite{yerokhin10a} to the leading order of $\alpha^3$ is adopted in this work,
\begin{widetext}
\begin{eqnarray}
H_{QED}^{(3)}\simeq \frac{4Z\alpha^3}{3}\left\{\dfrac{19}{30}+\ln\big[(Z\alpha)^{-2}\big]-\ln\big(\frac{k_0}{Z^2}\big)\right\}
              \big[ \delta^3({\bold{r}}_1)+\delta^3({\bold{r}}_2) \big] + O({\bold{r}}_{12})
               \,,\label{qed2}
\end{eqnarray}
\end{widetext}
where $\ln k_0$ is the Bethe logarithm, and $O({\bold{r}}_{12})$ represents remaining term connected with ${\bold{r}}_{12}$. We use $\ln k_0=$4.364 036 82(1) and 4.366 412 72(7)~\cite{drake99b} for the $2\,^3S_1$ and $2\,^1S_0$ states, respectively. Since $O({\bold{r}}_{12})$ contributes $-5.2\times 10^{-8}$ a.u. and $-5.6\times 10^{-9}$ a.u. to the energies of $2\,^1S_0$ and $2\,^3S_1$ states of helium, respectively, which is three and four orders of magnitude smaller than $1.66\times 10^{-5}$ a.u. and $1.67\times 10^{-5}$ a.u. from the first term of Eq.(\ref{qed2}). So in this work, the QED correction to the magic wavelengths is calculated by omitting the $O({\bold{r}}_{12})$ contributions.

The nonrelativistic dynamic hyperpolarizability for $S$ state is
\begin{eqnarray}
\gamma_0(\omega)=\frac{128\pi^2}{3}\big[\frac{1}{9}\mathcal{T}(1,0,1,\omega)+\frac{2}{45}\mathcal{T}(1,2,1,\omega)\big] \,, \label{e8}
\end{eqnarray}
where $\mathcal{T}(L_a,L_b,L_c,\omega)$ is written as
\begin{widetext}
\begin{eqnarray}
\mathcal{T}(L_a,L_b,L_c,\omega) &=& \sum_{mnk}\frac{\langle
N_gL_g\|T_1\|mL_a\rangle \langle mL_a\|T_1\|nL_b\rangle \langle
nL_b\|T_1\|kL_c\rangle \langle kL_c\|T_1\|N_gL_g\rangle }
{(\Delta E_{mg}-\omega)\Delta E_{ng}(\Delta E_{kg}-\omega)}\nonumber \\
&-&\delta_{L_b,L_g}\sum_{m}
\frac{|\langle N_gL_g\|T_1\|mL_a\rangle|^2}
{(\Delta E_{mg}-\omega)}\sum_{k}
\frac{|\langle N_gL_g\|T_1\|kL_c\rangle|^2}
{(\Delta E_{kg}-\omega)^2} \,.
\label{e9}
\end{eqnarray}
\end{widetext}
Compared with the dynamic dipole polarizabilities, the accurate calculation of the dynamic hyperpolarizabilities is much more challenging, since the formula involves three summations over different intermediated states, which means the calculation of hyperpolarizability depends on the completeness of the energy spectrum of more intermediate states.

\begin{figure}[!htbp]
\includegraphics[width=0.49\textwidth]{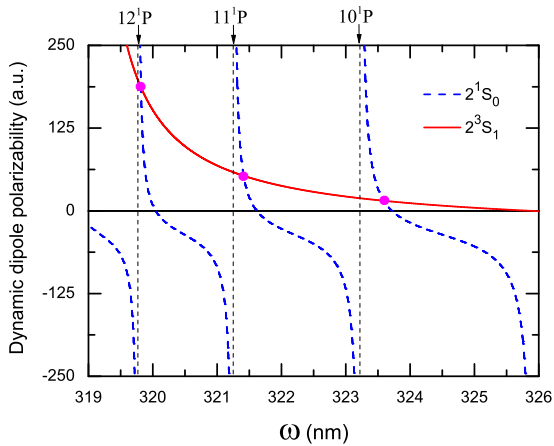}
\caption{\label{f1}(Color online) Dynamic dipole polarizabilities (in a.u.) of helium for the photon energy 319 nm $\leq\omega\leq $ 326 nm.
The solid red line and the dashed blue line represent the dynamic polarizabilities for $2\,^3S_1(M_J=\pm 1)$ and $2\,^1S_0$ states, respectively. The crossing points marked as solid magenta circle are the magic wavelengths. The vertical lines are the resonance transition positions, and the black line denotes a horizontal line. }
\end{figure}

The magic wavelengths of the $2\,^3S\rightarrow2\,^1S$ transition are determined from making the dynamic dipole polarizabilities of the $2\,^1S_0$ and $2\,^3S_1$ states equally. The accuracy of magic wavelengths depends on accurate energies and wavefunctions of initial and intermediate states. The high-precision B-spline RCI method was very successful in accurate calculation of atomic polarizabilities for the triplet $2\,^3S_1$ state of helium~\cite{zhang16a}. However, for the magic wavelengths around 320 nm of interest in the present work, it's clearly seen from Fig.~\ref{f1}, they are located at the edge of the $2\,^1S_0\rightarrow 10\,^1P_1$, $2\,^1S_0\rightarrow 11\,^1P_1$, and $2\,^1S_0\rightarrow 12\,^1P_1$ resonance transitions. The accurate determination of these magic wavelengths requires construction of sufficient configurations in an appropriate box size to make sure that all transition energies from the $2\,^1S_0$ state to the $10\,^1P_1$, $11\,^1P_1$, and $12\,^1P_1$ Rydberg states are accurate. This is a biggest challenge for our RCI calculation.

We optimize our RCI program by using OpenMP parallel and block calculations, which overcomes the problem of time consuming and large memory required in calculating the electron-electron Coulomb and Breit interaction integrals. Extensive tests of the numerical stability for energies, matrix elements, polarizabilities, and magic wavelengths of helium are carried out.

\begin{table}[!htbp]
\caption{\label{t1a} Energy (in a.u.) test of the $2\,^1S_0$ and $12\,^1P_1$ states of $^4$He. $N$ is the number of B-spline, $R_0$ is the box size, and $\ell_{max}$ presents the number of partial-wave. }
\begin{ruledtabular}
\begin{tabular}{clll}
&   \multicolumn{2}{c}{$2\,^1S_0$ } & \multicolumn{1}{c}{$12\,^1P_1$}  \\
 \cline{2-3} \cline{4-4}
\multicolumn{1}{l}{$\ell_{max}$} &   \multicolumn{1}{c}{$(N,R_0)$=(45,100) }  & \multicolumn{1}{c}{$(N,R_0)$=(50,600)} & \multicolumn{1}{c}{$(N,R_0)$=(50,600)} \\
   7  &  $-$2.145 783 68    & $-$2.145 782 91   &$-$2.003 297 046 \\
   8  &  $-$2.145 784 58    & $-$2.145 783 74   &$-$2.003 297 053 \\
   9  &  $-$2.145 785 15    & $-$2.145 784 25   &$-$2.003 297 057 \\
   10 &  $-$2.145 785 53    & $-$2.145 784 58   &$-$2.003 297 059 \\
   15 &  $-$2.145 786 27    & $-$2.145 785 21   &$-$2.003 297 063 \\
Extrap.& $-$2.145 786(1)    & $-$2.145 785(1)   &$-$2.003 297 06(1) \\
\end{tabular}
\end{ruledtabular}
\end{table}
\begin{table}[!htbp]
\caption{\label{t1} Comparison of present RCI energies (in a.u.) for some selective $n\,^1P_1\,(n \leq 13)$ states of $^4$He and $^3$He. The numbers in parentheses are computational uncertainties. }
\begin{ruledtabular}
\begin{tabular}{cll}
 \multicolumn{1}{c}{State}
&\multicolumn{2}{c}{$^4$He} \\
\cline{2-3}
&\multicolumn{1}{c}{RCI} &\multicolumn{1}{c}{Hylleraas~\cite{drake06a}}\\
\hline
$2\,^1P_1$  &$-$2.123 650 17(2)  &$-$2.123 654 51   \\
$3\,^1P_1$  &$-$2.054 968 56(2)  &$-$2.054 970 17   	\\
$4\,^1P_1$  &$-$2.030 896 59(2)  &$-$2.030 897 47   \\
$5\,^1P_1$  &$-$2.019 734 98(2)  &$-$2.019 735 59    \\
$6\,^1P_1$  &$-$2.013 664 02(2)  &$-$2.013 664 52    \\
$7\,^1P_1$  &$-$2.009 999 98(2)  &$-$2.010 000 41   	 \\
$8\,^1P_1$  &$-$2.007 620 18(2)  &$-$2.007 620 57    \\
$9\,^1P_1$  &$-$2.005 987 70(2)  &$-$2.005 988 07    \\
$10\,^1P_1$ &$-$2.004 819 48(2)  &$-$2.004 819 84   	\\
$11\,^1P_1$ &$-$2.003 954 83(2)  &                  \\
$12\,^1P_1$ &$-$2.003 297 06(1)  &                  	 \\
$13\,^1P_1$ &$-$2.002 784 91(4)  &                  	\\
\hline
\hline
 \multicolumn{1}{c}{State}
&\multicolumn{2}{c}{$^3$He}\\
\cline{2-3}
&\multicolumn{1}{c}{RCI} &\multicolumn{1}{c}{Hylleraas~\cite{drake06a}}\\
\hline
$2\,^1P_1$  &	$-$2.123 552 83(2)	&	$-$2.123 557 20	\\
$3\,^1P_1$  &	$-$2.054 875 71(2)	&	$-$2.054 877 36	\\
$4\,^1P_1$  &	$-$2.030 805 19(2)	&	$-$2.030 806 11 \\
$5\,^1P_1$  &	$-$2.019 644 21(2)	&	$-$2.019 644 86 \\
$6\,^1P_1$  &	$-$2.013 573 59(2)	&	$-$2.013 574 20 \\
$7\,^1P_1$  &	$-$2.009 909 74(2)	&	$-$2.009 910 20	 \\
$8\,^1P_1$  &	$-$2.007 530 06(2)	&	$-$2.007 530 50 \\
$9\,^1P_1$  &	$-$2.005 897 67(2)	&	$-$2.005 898 08 \\
$10\,^1P_1$ &	$-$2.004 729 51(2)	&	$-$2.004 729 91	\\
$11\,^1P_1$ &	$-$2.003 864 90(2)	&	\\
$12\,^1P_1$ &	$-$2.003 207 09(4)   &	 \\
$13\,^1P_1$ &	$-$2.002 695 04(4)	&	\\
\end{tabular}
\end{ruledtabular}
\end{table}

In order to get accurate energies for the intermediate $n\,^1P_1\,(n=10\sim13)$ states, we fix the box size $R_0=600$ a.u. and increase the numbers of partial-wave $l_{max}$, and B-spline basis sets $N$ to test the convergence of energies. Seen from the Table~\ref{t1a}, when fixing $N=50$ to increase $\ell_{max}$, we get the converged energy of $-$2.003 297 06(1) a.u. for the $12\,^1P_1$ state, which has eight significant digits.

For other $n\,^1P_1$ states, seen from the Table~\ref{t1}, all the energies for $n\,^1P_1 (n\leq 13)$ intermediate states have 8 significant digits. That means the energy accuracy for all the states, even for the Rydberg states, can be guaranteed to the same level of precision from one diagonalization in present RCI calculations. The Hylleraas energies~\cite{drake06a} of Table~\ref{t1} are derived by combining the values in the Tables 11.7 and 11.8 of Ref.~\cite{drake06a} and the ground-state energy of $-1.999\,832\,572\, 508$ a.u. of He$^+$. Compared with the Hylleraas energies~\cite{drake06a}, which include the finite nuclear mass, relativistic, and anomalous magnetic moment corrections, our RCI energies are in good agreement with Hylleraas energies~\cite{drake06a}.

But for the energy of $2\,^1S_0$ state, seen from the Table~\ref{t1a}, since the electron-electron correlation is much larger than the $12\,^1P_1$ state, we cannot get 8 significant digits from present largest-scale RCI calculation. Even we decrease the box size $R_0$ to 100 a.u. and fix $N=45$ to increase $\ell_{max}=15$, the convergent energy is $-$2.145 786(1) a.u., which is less accurate than the $n\,^1P_1 (n\leq13)$ states by one order of magnitude, and just has seven same digits compared with the best value of $-$2.145 786 909 a.u., ~\cite{yerokhin10a}. So in the later determination of magic wavelengths, we replace our RCI energies of the $2\,^1S_0$ state of helium with the values of Ref.~\cite{yerokhin10a}.

\begin{table}[!htbp]
\caption{\label{t2} Comparison of some reduced matrix elements between present RCI calculations and Hylleraas calculations~\cite{drake07a} of the $2\,^1S_0\rightarrow n\,^1P_1 (n \leq 13)$ transitions for $^4$He and $^3$He. The numbers in parentheses are computational uncertainties. }
\begin{ruledtabular}
\begin{tabular}{cllc}
 \multicolumn{1}{c}{$2\,^1S_0\rightarrow n\,^1P_1$}
&\multicolumn{2}{c}{RCI}&\multicolumn{1}{c}{Hylleraas~\cite{drake07a}}\\
\cline{2-3}
&\multicolumn{1}{c}{$^3$He} &\multicolumn{1}{c}{$^4$He}&\multicolumn{1}{c}{$^4$He} \\
\hline
 2   &5.052 46(2) &5.052 06(8)  &5.050 977  \\
 3   &1.580 76(2) &1.580 81(2)  &1.581 082  \\
 4   &0.801 02(2) &0.801 03(2)  &0.801 106  \\
 5   &0.515 54(2) &0.515 53(2)  &0.515 578  \\
 6   &0.371 14(2) &0.371 14(2)  &0.371 159  \\
 7   &0.285 12(2) &0.285 12(2)  &0.285 131  \\
 8   &0.228 58(2) &0.228 58(2)  &0.228 590  \\
 9   &0.188 89(2) &0.188 89(2)  &0.188 899 \\
 10  &0.159 68(2) &0.159 67(2)  &0.159 686 \\
 11  &0.137 40(2) &0.137 39(2)  &   \\
 12  &0.119 92(2) &0.119 92(2)  &  \\
 13  &0.105 90(2) &0.105 89(2)  & \\
\end{tabular}
\end{ruledtabular}
\end{table}

Table~\ref{t2} gives a comparison of the reduced matrix elements for the dipole allowed $2\,^1S_0\rightarrow n\,^1P_1$($n\leq13$) transitions. The Hylleraas values~\cite{drake07a} includes the finite nuclear mass and the leading relativistic corrections. Compared with the Hylleraas values, present RCI results have four same digits for the $2\,^1S_0\rightarrow n\,^1P_1$($n\geq 5$) transitions. The energies of $n\,^1S_0$, $n\,^3S_1$ and $n\,^3P_J$ states, and the reduced matrix elements for $2\,^1S_0\rightarrow n\,^3P_1$, $2\,^3S_1\rightarrow n\,^3P_J$, and $2\,^3S_1\rightarrow n\,^1P_1$ transitions with $n\leq13$ for $^3$He and $^4$He are presented in Supplemental Material~\cite{supp18a}.

\begin{table}[!htbp]
\caption{\label{t3} Convergence test of 319.8 nm magic wavelength for the $2\,^3S_1(M_{J}=\pm1)\rightarrow2\,^1S_0$ transition of $^4$He as the number of B-splines, $N$ increased with the number of the partial wave, $\ell_{max}=7$, and as $\ell_{max}$ increased with $N=50$ by choosing the box size, $R_0=$ 600 a.u. The convergence test for the dynamic dipole polarizabilities at the corresponding magic wavelengths are also listed.}
\begin{ruledtabular}
\begin{tabular}{lll}
&\multicolumn{2}{c}{$\ell_{max}$=7}\\
\cline{2-3}
\multicolumn{1}{l}{$N$}&\multicolumn{1}{c}{$\lambda_m$} &\multicolumn{1}{c}{$\alpha_1(\omega)$}  \\
\hline
40             &319.828 217      &183.903 70    \\
45             &319.815 649      &186.660 50    \\
50             &319.814 254      &186.967 81     \\	
55             &319.814 140      &186.992 26    \\	
60             &319.814 128      &186.995 32    \\
\multicolumn{1}{l}{$\ell_{max}$}&\multicolumn{2}{c}{$N$=50}\\
\cline{2-3}
10             &319.814 287     &186.959 58 \\
15             &319.814 299     &186.957 01 \\
20             &319.814 300     &186.956 61 \\
\hline
Extrap.        &319.814 3(4)    &186.96(6) \\
\end{tabular}
\end{ruledtabular}
\end{table}

Since the 319.8 nm magic wavelength was proposed to trap helium for high-precision measurement, Table~\ref{t3} lists the convergent test for this particular magic wavelength of the $2\,^3S_1(M_{J}=\pm1)\rightarrow2\,^1S_0$ transition of $^4$He. The corresponding dynamic dipole polarizabilities at the magic wavelengths are also listed. It is seen that both the parameters, $N$ and $\ell_{max}$, affect the convergent rate of numerical values. According to the values in the last three lines, we can obtain the extrapolated value of 319.814 3(4) nm for the magic wavelength. In order to take account of the incompleteness of configurations, the uncertainty of 319.814 3(4) nm is obtained by doubling the difference of 319.814 300 nm and 319.814 128 nm for the sake of conservativeness. Similarly, we can get the extrapolated polarizability of 186.96(6) a.u., which is more accurate than the semi-empirical result of 189.3 a.u.~\cite{notermans14b}.

\begin{table}[!htbp]
\caption{\label{t3b} Convergence test of 319.8 nm magic wavelength for the $2\,^3S_1(M_{J}=\pm1)\rightarrow2\,^1S_0$ transition and the $2\,^3S_1 (F=\frac{3}{2},M_{F}=\frac{3}{2})\rightarrow2\,^1S_0 (F=\frac{1}{2}, M_{F}=\frac{1}{2})$ hyperfine transition of $^3$He as the number of B-splines, $N$, increased by fixing the number of the partial wave as $\ell_{max}=7$ and choosing the box size as $R_0=$ 600 a.u. The dynamic dipole polarizabilities at the corresponding magic wavelengths are also listed. }
\begin{ruledtabular}
\begin{tabular}{llllll}

&\multicolumn{2}{l}{$2\,^3S_1(M_{J}=\pm1)$} &&\multicolumn{2}{l}{$2\,^3S_1 (F=\frac{3}{2},M_{F}=\frac{3}{2})$}  \\
&\multicolumn{2}{l}{$\rightarrow2\,^1S_0$ } &&\multicolumn{2}{l}{$\rightarrow2\,^1S_0 (F=\frac{1}{2}, M_{F}=\frac{1}{2})$}  \\
\cline{2-3} \cline{5-6}
\multicolumn{1}{l}{$N$}&\multicolumn{1}{c}{$\lambda_m$} &\multicolumn{1}{c}{$\alpha_1(\omega)$} &&\multicolumn{1}{c}{$\lambda_m$} &\multicolumn{1}{c}{$\alpha_1(\omega)$}\\
\hline
40                      &319.843 338    &184.136 47  &&319.843 372  &183.888 81 \\
45                      &319.830 791    &186.897 83  &&319.830 832  &186.641 95 \\
50                      &319.829 394    &187.201 98  &&319.829 437  &186.945 04  \\	
55                      &319.829 278    &187.237 65  &&319.829 320  &186.980 95  \\	
60                      &319.829 266    &187.232 93  &&319.829 308  &186.976 20 \\
Extrap.                 &319.829 2(4)   &187.22(6)   &&319.829 3(4) &186.96(6) \\
\end{tabular}
\end{ruledtabular}
\end{table}

For the $^3$He atom, In the Table~\ref{t3b}, the convergence test of the 319.8 nm magic wavelength with and without the hyperfine effect are presented. For the $2\,^3S_1(M_{J}=\pm1)\rightarrow2\,^1S_0$ transition, the extrapolated values of 319.829 2(4) nm and 187.22(6) a.u. are, respectively, for the magic wavelength and the corresponding dynamic dipole polarizability. For the $2\,^3S_1 (F=\frac{3}{2},M_{F}=\frac{3}{2})\rightarrow2\,^1S_0 (F=\frac{1}{2}, M_{F}=\frac{1}{2})$ hyperfine transition, we use the hyperfine energy shifts of Ref.~\cite{morton06b} for the $2\,^{1}S_0$, $2\,^{3}S_1$, and $n\,^{1,3}P_J\, (n\leq 10)$ states. For higher $n\,^{1,3}P_J$ intermediate states, the hyperfine energy shifts are obtained by fitting the hyperfine splitting of $n\,^{1,3}P_J\, (n\leq 10)$ states. The reduced matrix elements between hyperfine levels can be transformed by the Eq.(4) of Ref.~\cite{jiang13c}. Then we replace the hyperfine energies and matrix elements into the Eqs.(\ref{e4})-(\ref{e7}) to get dynamic dipole polarizabilities for extracting the magic wavelengths. We find that the hyperfine effect has large correction to $\alpha_1(\omega)$, but only increase about 0.1 picometer (pm) on the extrapolated $\lambda_m$ of 319.829 2(4) nm, which can be taken as one source of the uncertainty in the final recommended magic wavelength.

\begin{table*}
\caption{\label{t4} The first nine magic wavelengths (in nm) of $2\,^3S_1\rightarrow2\,^1S_0$ transition of helium. The numbers in parentheses are computational uncertainties. The uncertainties in present RCI values evaluated from the incompleteness of configuration space. The values of QED correction only represents the convergence results of our numerical calculation, the uncertainty would be multiplied by 10 in the final QED correction for conservativeness. }
\begin{ruledtabular}
\begin{tabular}{lcllllllllllll}
&\multicolumn{1}{c}{Hyllerass-B-splines}
& &\multicolumn{5}{c}{RCI }  & &\multicolumn{1}{c}{QED}&&\multicolumn{2}{c}{Ref.~\cite{notermans14b}}\\
\cline{2-2} \cline{4-8} \cline{10-10} \cline{12-13}
&\multicolumn{1}{c}{$2\,^3S\rightarrow2\,^1S$ }
& &\multicolumn{2}{c}{$2\,^3S_1(M_{J}=\pm1)\rightarrow2\,^1S_0$} & &\multicolumn{2}{c}{$2\,^3S_1(M_{J}=0)\rightarrow2\,^1S_0$} & &\multicolumn{1}{c}{$2\,^3S_1\rightarrow2\,^1S_{0}$} & &\multicolumn{2}{c}{$2\,^3S_1\rightarrow2\,^1S_{0}$}\\
\cline{2-2}  \cline{4-5} \cline{7-8}\cline{10-10}\cline{12-13}
\multicolumn{1}{l}{No.}&\multicolumn{1}{c}{$^\infty$He}
& &\multicolumn{1}{l}{$^4$He}&\multicolumn{1}{l}{$^3$He} & &\multicolumn{1}{l}{$^4$He}&\multicolumn{1}{l}{$^3$He} & && &\multicolumn{1}{l}{$^4$He}&\multicolumn{1}{l}{$^3$He}\\
\hline
1   &412.16(2)	  & &	412.167(5)    &412.166(5)    &&412.173(8)   &412.177(7)    && $-$0.000 43(2) &&411.863    & 	\\
2	&352.299(4)	  & &	352.335(3)    &352.351(4)    &&352.336(3)   &352.352(4)    && 0.001 12(2) &&352.242    & 	\\
3	&338.641(2)   & &	338.681 7(5)  &338.697 2(5)  &&338.681 8(5) &338.697 4(5)  && 0.001 07(2) &&338.644    &	 \\
4	&331.240(1)   & &	331.282 7(4)  &331.298 1(4)  &&331.282 8(4) &331.298 3(4)  && 0.001 04(2) &&331.268    &	\\
5	&326.633(1)   & &	326.677 0(4)  &326.692 2(4)  &&326.677 1(4) &326.692 3(4)  && 0.001 03(2) &&326.672    &	 \\
6	&323.544(1)   & &	323.587 9(4)  &323.603 1(4)  &&323.588 0(4) &323.603 4(4)  && 0.001 02(2) &&323.587    &323.602	 \\
7	&321.366(1)   & &	321.409 5(4)  &321.424 7(4)  &&321.409 6(4) &321.424 9(4)  && 0.001 01(2) &&321.409    &321.423	\\
8	&319.771(1)   & &	319.814 3(4)  &319.829 2(4)  &&319.814 4(4) &319.829 4(4)  && 0.001 00(2) &&319.815    &319.830	\\
9	&318.567(1)   & &	318.610 5(8)  &318.625 6(8)  &&318.610 6(8) &318.625 8(8)  && 0.001 11(2) &&318.611    &318.626 	\\
\end{tabular}
\end{ruledtabular}
\end{table*}

In addition, the QED correction to all the magic wavelengths are extracted by performing the calculation of QED correction to the dynamic dipole polarizabilities. The values are listed in the Table~\ref{t4}. Especially for the 319.8 nm magic wavelength, leading order of QED correction is 0.001 00(2) nm. Other terms, such as the second derivative of the Bethe logarithm, Araki-Sucher term, and high-order QED corrections, would bring possible sources of the error. For the sake of conservativeness, we can multiply the uncertainty by 10. So the final QED correction of 0.0010(2) nm is indicated to the 319.8 nm magic wavelength, which can be added into present RCI values, then we can get the recommended magic wavelengths of 319.815 3(6) nm and 319.815 4(6) nm, respectively, for the $2\,^3S_1(M_{J}=\pm1)\rightarrow2\,^1S_0$ and $2\,^3S_1(M_{J}=0)\rightarrow2\,^1S_0$ transitions of $^4$He. Similarly, with the QED and hyperfine structure corrections taken into account for $^3$He, we can give the recommended values of 319.830 2(7) nm and 319.830 4(7) nm, respectively,  for the $2\,^3S_1(M_{J}=\pm1)\rightarrow2\,^1S_0$ and $2\,^3S_1(M_{J}=0)\rightarrow2\,^1S_0$ transitions of $^3$He. Our recommended value of 319.815 3(6) nm for $\,^4$He agrees well with recent measurement result of 319.815 92(15) nm~\cite{wim18a}. And present magic wavelength of $^3$He would provide theoretical reference for designing ODT experiment to help resolving the nuclear radius discrepancy of helium isotopes.

Except the important application of 319.8 nm magic wavelength of helium, both of the 321.4 nm and 323.5 nm magic wavelengths can also be used to design experiments once high-power laser can be realized. The magic wavelengths obtained from present RCI calculations are 323.587 9(4) nm and 321.409 5(4) nm for the $2\,^3S_1(M_{J}=\pm1)\rightarrow2\,^1S_0$ transition of $^4$He. Taking the QED correction into account, we recommend 323.588 9(6) nm and 321.410 5(6) nm as the final values of magic wavelengths. Similarly, for the $2\,^3S_1(M_{J}=\pm1)\rightarrow2\,^1S_0$ transition of $^3$He, with the QED and hyperfine corrections included, the magic wavelengths of 323.604 1(7) nm and 321.425 7(7) nm are recommended.

Table~\ref{t4} summarizes the first nine magic wavelengths in the range of 318 - 413 nm from two independent calculations of the Hyllerass-B-splines and RCI methods. All the values from two different theoretical methods are consistent. The relativistic and finite nuclear mass corrections on all the magic wavelengths are less than 60 pm. For the RCI calculation, the difference of all the magic wavelengths between $^4$He and $^3$He are less than 17 pm. It's noticed that the QED corrections listed in the Table~\ref{t4} only represent the convergence results of present numerical calculation, the uncertainty may be multiplied by 10 in the final QED correction for conservatively taking other neglected contributions into account.

\begin{table}
\caption{\label{t5} Dynamic hyperpolarizabilities (in a.u.) at the nine magic wavelengths for the $2\,^1S$ and $2\,^3S$ states of $^\infty$He. The high-order Ac Stark shifts (in Hz) are also presented in the last column. The electric field intensity $F \approx 1.58\times 10^{-7}$ a.u. is evaluated from real experimental condition~\cite{wim18a}. The numbers in the square brackets denote powers of ten. The numbers in parentheses are computational uncertainties.}
\begin{ruledtabular}
\begin{tabular}{llll}
  \multicolumn{1}{c}{No.} &\multicolumn{1}{c}{$2\,^1S$ } & \multicolumn{1}{c}{$2\,^3S$ } & \multicolumn{1}{c}{$\frac{1}{24}\Delta\gamma_0(\omega)F^4 $}\\ \hline
   1     &$-$6(4)[6]      &$-$2.3(5)[8]  &$-$3.9[-5] \\
   2     &$-$1.2(1)[6]    &$-$3.6(9)[7]  &$-$6.0[-6] \\
   3     &$-$4.1(9)[6]    &$-$2.4(9)[7]  &$-$3.4[-6] \\
   4     &$-$1.5(7)[7]    &$-$3.7(8)[7]  &$-$3.6[-6] \\
   5     &$-$2.9(4)[7]    &$-$5.7(6)[7]  &$-$4.8[-6] \\
   6     &$-$9.6(2)[7]    &$-$1.1(3)[8]  &$-$2.6[-6]\\
   7     &$-$5.4(1)[8]    &$-$3.7(3)[8]  &3.1[-5]\\
   8     &$-$1.0(1)[10]   &$-$3.4(1)[9]  &1.2[-3]\\
   9     &4.2(1)[11]      &3.9(1)[10]    &$-$6.6[-2] \\
\end{tabular}
\end{ruledtabular}
\end{table}

Table~\ref{t5} presents the dynamic hyperpolarizabilities at the nine magic wavelengths of Table~\ref{t4} for the $2\,^1S$ and $2\,^3S$ states of $^\infty$He by using the Hyllerass-B-splines method. And the high-order Ac Stark shift at each magic wavelength is also estimated. Especially, for the 319.8 nm magic wavelength, the dynamic hyperpolarizabilities are $-1.0\times 10^{10}$ a.u. and $-3.4\times 10^9$ a.u. for the $2\,^1S$ and $2\,^3S$ states of $^\infty$He, respectively. The difference of the dynamic hyperpolarizabilities for the $2\,^3S\rightarrow2\,^1S$ transition is $\Delta\gamma_0(\omega)=6.7\times 10^9$ a.u. If the power of the incident trapping laser beam is $P=0.2\,W$ with beam waist $w_0=85\,\mu m$, then we can get the electric field intensity $F \approx 1.58\times 10^{-7}$ a.u. Accordingly, the higher-order Ac Stark shift is evaluated as $\frac{1}{24}\Delta\gamma_0(\omega)F^4\approx 1.7\times 10^{-19}$ a.u.$\approx 1.2$ mHz, it is smaller by six orders of magnitude than the 1.8 kHz uncertainty of the absolute frequency for the $2\,^3S_1\rightarrow2\,^1S_0$ transition of $^4$He~\cite{rooij11a}, which indicates the high-order Ac Stark shift can be neglected for the precision spectroscopy measurement of the $2\,^3S_1\rightarrow2\,^1S_0$ transition of helium by implementation of a magic wavelength trap.

In summary, the improved RCI method enables us to calculate the dynamic dipole polarizabilities in wide range of laser frequency for both $2\,^3S_1$ and $2\,^1S_0$ states of helium. A series of magic wavelengths for $2\,^3S_1\rightarrow2\,^1S_0$ forbidden transition of $^4$He and $^3$He are accurately determined. The non-relativistic calculations of magic wavelength for $^\infty$He are also carried out by using the Hylleraas-B-spline method. Further, the leading order of QED corrections on the magic wavelengths have taken into account. For $^3$He, the correction from hyperfine structure to the magic wavelengths has been calculated. In addition, the high-order Ac Stark shift related with the dynamic hyperpolarizabilities are estimated. All the magic wavelengths from two different theoretical methods are consistent. Present recommended magic wavelength of 319.815 3(6) nm for $^4$He is in good agreement with the high-precision measurement value of 319.815 92(15) nm~\cite{wim18a,wim18b}. Present magic wavelength of 319.830 2(7) nm for $^3$He provides important reference for experimental design of a magic wavelength trap to eliminate the Ac Stark shift for the precision spectroscopy of the $2\,^3S_1\rightarrow2\,^1S_0$ transition of helium in future.

We thank Wim Vassen for early suggestion to carry out this work. We also thank Yong-Bo Tang for his discussion for the improvement of present RCI program. This work was supported by the Strategic Priority Research Program of the Chinese Academy of Sciences, Grant Nos.XDB21010400 and XDB21030300, by the National Key Research and Development Program of China under Grant No.2017YFA0304402, and by the National Natural Science Foundation of China under Grants Nos.11474319, 11704398, 11774386, 91536102 and 11674253.


\begin{thebibliography}{36}
\expandafter\ifx\csname natexlab\endcsname\relax\def\natexlab#1{#1}\fi
\expandafter\ifx\csname bibnamefont\endcsname\relax
  \def\bibnamefont#1{#1}\fi
\expandafter\ifx\csname bibfnamefont\endcsname\relax
  \def\bibfnamefont#1{#1}\fi
\expandafter\ifx\csname citenamefont\endcsname\relax
  \def\citenamefont#1{#1}\fi
\expandafter\ifx\csname url\endcsname\relax
  \def\url#1{\texttt{#1}}\fi
\expandafter\ifx\csname urlprefix\endcsname\relax\def\urlprefix{URL }\fi
\providecommand{\bibinfo}[2]{#2}
\providecommand{\eprint}[2][]{\url{#2}}

\bibitem[{\citenamefont{Pohl et~al.}(2010)}]{pohl10a}
\bibinfo{author}{\bibfnamefont{R.}~\bibnamefont{Pohl}}, 
\bibinfo{author}{\bibfnamefont{A.}~\bibnamefont{Antognini}}, 
\bibinfo{author}{\bibfnamefont{F.}~\bibnamefont{Nez}}, 
\bibinfo{author}{\bibfnamefont{F. D.}~\bibnamefont{Amaro}},
\bibinfo{author}{\bibfnamefont{F.}~\bibnamefont{Biraben}}, 
\bibinfo{author}{\bibfnamefont{J. M. R.}~\bibnamefont{Cardoso}},
\bibinfo{author}{\bibfnamefont{D. S.}~\bibnamefont{Covita}},
\bibinfo{author}{\bibfnamefont{A.}~\bibnamefont{Dax}},
\bibinfo{author}{\bibfnamefont{S.}~\bibnamefont{Dhawan}},
\bibinfo{author}{\bibfnamefont{L. M. P.}~\bibnamefont{Fernandes}},
\bibnamefont{et~al.}, \bibinfo{journal}{Nature} \textbf{\bibinfo{volume}{466}},
\bibinfo{pages}{213} (\bibinfo{year}{2010}).

\bibitem[{\citenamefont{Antognini et~al.}(2013)}]{antognini13a}
\bibinfo{author}{\bibfnamefont{A.}~\bibnamefont{Antognini}},
\bibinfo{author}{\bibfnamefont{F.}~\bibnamefont{Nez}},
\bibinfo{author}{\bibfnamefont{K.}~\bibnamefont{Schuhmann}},
\bibinfo{author}{\bibfnamefont{F. D.}~\bibnamefont{Amaro}},
\bibinfo{author}{\bibfnamefont{F.}~\bibnamefont{Biraben}},
\bibinfo{author}{\bibfnamefont{J. M. R.}~\bibnamefont{Cardoso}},
\bibinfo{author}{\bibfnamefont{D. S.}~\bibnamefont{Covita}},
\bibinfo{author}{\bibfnamefont{A.}~\bibnamefont{Dax}},
\bibinfo{author}{\bibfnamefont{S.}~\bibnamefont{Dhawan}},
\bibinfo{author}{\bibfnamefont{M.}~\bibnamefont{Diepold}},
 \bibnamefont{et~al.}, \bibinfo{journal}{Science}
  \textbf{\bibinfo{volume}{339}}, \bibinfo{pages}{417} (\bibinfo{year}{2013}).

\bibitem[{\citenamefont{Mohr et~al.}(2016)\citenamefont{Mohr, Newell, and
  Taylor}}]{mohr16a}
\bibinfo{author}{\bibfnamefont{P.~J.} \bibnamefont{Mohr}},
  \bibinfo{author}{\bibfnamefont{D.~B.} \bibnamefont{Newell}},
  \bibnamefont{and} \bibinfo{author}{\bibfnamefont{B.~N.}
  \bibnamefont{Taylor}}, \bibinfo{journal}{Rev. Mod. Phys.}
  \textbf{\bibinfo{volume}{88}}, \bibinfo{pages}{035009}
  (\bibinfo{year}{2016}).

\bibitem[{\citenamefont{Beyer et~al.}(2017)}]{beyer17a}
\bibinfo{author}{\bibfnamefont{A.}~\bibnamefont{Beyer}}, 
\bibinfo{author}{\bibfnamefont{L.}~\bibnamefont{Maisenbacher}}, 
\bibinfo{author}{\bibfnamefont{A.}~\bibnamefont{Matveev}}, 
\bibinfo{author}{\bibfnamefont{R.}~\bibnamefont{Pohl}}, 
\bibinfo{author}{\bibfnamefont{K.}~\bibnamefont{Khabarova}}, 
\bibinfo{author}{\bibfnamefont{A.}~\bibnamefont{Grinin}},
\bibinfo{author}{\bibfnamefont{D. C.}~\bibnamefont{Yost}}, 
\bibinfo{author}{\bibfnamefont{T. W.}~\bibnamefont{H{\"a}nsch}},
\bibinfo{author}{\bibfnamefont{N.}~\bibnamefont{Kolachevsky}},
 \bibnamefont{and} \bibinfo{author}{\bibfnamefont{T.}~\bibnamefont{Udem}},
  \bibinfo{journal}{Science} \textbf{\bibinfo{volume}{358}},
  \bibinfo{pages}{79} (\bibinfo{year}{2017}).

\bibitem[{\citenamefont{Tan et~al.}(2011)\citenamefont{Tan, Brewer, and
  Guise}}]{tan11a}
\bibinfo{author}{\bibfnamefont{J.~N.} \bibnamefont{Tan}},
  \bibinfo{author}{\bibfnamefont{S.~M.} \bibnamefont{Brewer}},
  \bibnamefont{and} \bibinfo{author}{\bibfnamefont{N.~D.} \bibnamefont{Guise}},
  \bibinfo{journal}{Phys. Scr.} \textbf{\bibinfo{volume}{2011}},
  \bibinfo{pages}{014009} (\bibinfo{year}{2011}).

\bibitem[{\citenamefont{Antognini et~al.}(2011)}]{antognini11a}
\bibinfo{author}{\bibfnamefont{A.}~\bibnamefont{Antognini}},
\bibinfo{author}{\bibfnamefont{F.}~\bibnamefont{Biraben}},
\bibinfo{author}{\bibfnamefont{J. M. R.}~\bibnamefont{Cardoso}},
\bibinfo{author}{\bibfnamefont{D. S.}~\bibnamefont{Covita}},
\bibinfo{author}{\bibfnamefont{A.}~\bibnamefont{Dax}},
\bibinfo{author}{\bibfnamefont{L. M. P.}~\bibnamefont{Fernandes}},
\bibinfo{author}{\bibfnamefont{A. L.}~\bibnamefont{Gouvea}},
\bibinfo{author}{\bibfnamefont{T. W.}~\bibnamefont{H?nsch}},
\bibinfo{author}{\bibfnamefont{M.}~\bibnamefont{Hildebrandt}},
\bibnamefont{and}
\bibinfo{author}{\bibfnamefont{P.}~\bibnamefont{Indelicato}},
 \bibinfo{journal}{Can. J. Phys.}
  \textbf{\bibinfo{volume}{89}}, \bibinfo{pages}{47} (\bibinfo{year}{2011}).

\bibitem[{\citenamefont{{van Rooij} et~al.}(2011)\citenamefont{{van Rooij},
  {Borbely}, {Simonet}, {Hoogerland}, {Eikema}, {Rozendaal}, and
  {Vassen}}}]{rooij11a}
\bibinfo{author}{\bibfnamefont{R.}~\bibnamefont{{van Rooij}}},
  \bibinfo{author}{\bibfnamefont{J.~S.} \bibnamefont{{Borbely}}},
  \bibinfo{author}{\bibfnamefont{J.}~\bibnamefont{{Simonet}}},
  \bibinfo{author}{\bibfnamefont{M.~D.} \bibnamefont{{Hoogerland}}},
  \bibinfo{author}{\bibfnamefont{K.~S.~E.} \bibnamefont{{Eikema}}},
  \bibinfo{author}{\bibfnamefont{R.~A.} \bibnamefont{{Rozendaal}}},
  \bibnamefont{and} \bibinfo{author}{\bibfnamefont{W.}~\bibnamefont{{Vassen}}},
  \bibinfo{journal}{Science} \textbf{\bibinfo{volume}{333}},
  \bibinfo{pages}{196} (\bibinfo{year}{2011}).

\bibitem[{\citenamefont{{van Leeuwen} and Vassen}(2006)}]{leeuwen06a}
\bibinfo{author}{\bibfnamefont{K.~A.~H.} \bibnamefont{{van Leeuwen}}}
  \bibnamefont{and} \bibinfo{author}{\bibfnamefont{W.}~\bibnamefont{Vassen}},
  \bibinfo{journal}{Europhys. Lett.} \textbf{\bibinfo{volume}{76}},
  \bibinfo{pages}{409} (\bibinfo{year}{2006}).

\bibitem[{\citenamefont{Shiner et~al.}(1995)\citenamefont{Shiner, Dixson, and
  Vedantham}}]{shiner95a}
\bibinfo{author}{\bibfnamefont{D.}~\bibnamefont{Shiner}},
  \bibinfo{author}{\bibfnamefont{R.}~\bibnamefont{Dixson}}, \bibnamefont{and}
  \bibinfo{author}{\bibfnamefont{V.}~\bibnamefont{Vedantham}},
  \bibinfo{journal}{Phys. Rev. Lett.} \textbf{\bibinfo{volume}{74}},
  \bibinfo{pages}{3553} (\bibinfo{year}{1995}).

\bibitem[{\citenamefont{Morton et~al.}(2006{\natexlab{a}})\citenamefont{Morton,
  Wu, and Drake}}]{morton06a}
\bibinfo{author}{\bibfnamefont{D.~C.} \bibnamefont{Morton}},
  \bibinfo{author}{\bibfnamefont{Q.}~\bibnamefont{Wu}}, \bibnamefont{and}
  \bibinfo{author}{\bibfnamefont{G.~W.~F.} \bibnamefont{Drake}},
  \bibinfo{journal}{Phys. Rev. A} \textbf{\bibinfo{volume}{73}},
  \bibinfo{pages}{034502} (\bibinfo{year}{2006}{\natexlab{a}}).

\bibitem[{\citenamefont{Pastor et~al.}(2004)\citenamefont{Pastor, Giusfredi,
  Natale, Hagel, de~Mauro, and Inguscio}}]{pastor04a}
\bibinfo{author}{\bibfnamefont{P.~C.} \bibnamefont{Pastor}},
  \bibinfo{author}{\bibfnamefont{G.}~\bibnamefont{Giusfredi}},
  \bibinfo{author}{\bibfnamefont{P.} \bibnamefont{DeNatale}},
  \bibinfo{author}{\bibfnamefont{G.}~\bibnamefont{Hagel}},
  \bibinfo{author}{\bibfnamefont{C.}~\bibnamefont{de~Mauro}}, \bibnamefont{and}
  \bibinfo{author}{\bibfnamefont{M.}~\bibnamefont{Inguscio}},
  \bibinfo{journal}{Phys. Rev. Lett.} \textbf{\bibinfo{volume}{92}},
  \bibinfo{pages}{023001} (\bibinfo{year}{2004}).

\bibitem[{\citenamefont{Pastor et~al.}(2012)\citenamefont{Pastor, Consolino,
  Giusfredi, De~Natale, Inguscio, Yerokhin, and Pachucki}}]{pastor12a}
\bibinfo{author}{\bibfnamefont{P.~C.} \bibnamefont{Pastor}},
  \bibinfo{author}{\bibfnamefont{L.}~\bibnamefont{Consolino}},
  \bibinfo{author}{\bibfnamefont{G.}~\bibnamefont{Giusfredi}},
  \bibinfo{author}{\bibfnamefont{P.}~\bibnamefont{De~Natale}},
  \bibinfo{author}{\bibfnamefont{M.}~\bibnamefont{Inguscio}},
  \bibinfo{author}{\bibfnamefont{V.~A.} \bibnamefont{Yerokhin}},
  \bibnamefont{and} \bibinfo{author}{\bibfnamefont{K.}~\bibnamefont{Pachucki}},
  \bibinfo{journal}{Phys. Rev. Lett.} \textbf{\bibinfo{volume}{108}},
  \bibinfo{pages}{143001} (\bibinfo{year}{2012}).

\bibitem[{\citenamefont{Pachucki and Yerokhin}(2015)}]{pachucki15a}
\bibinfo{author}{\bibfnamefont{K.}~\bibnamefont{Pachucki}} \bibnamefont{and}
  \bibinfo{author}{\bibfnamefont{V.~A.} \bibnamefont{Yerokhin}},
  \bibinfo{journal}{J. Phys. Chem. Ref. Data} \textbf{\bibinfo{volume}{44}},
  \bibinfo{pages}{031206} (\bibinfo{year}{2015}).

\bibitem[{\citenamefont{Patk\'o\ifmmode~\check{s}\else \v{s}\fi{}
  et~al.}(2016)\citenamefont{Patk\'o\ifmmode~\check{s}\else \v{s}\fi{},
  Yerokhin, and Pachucki}}]{patkos16a}
\bibinfo{author}{\bibfnamefont{V.}~\bibnamefont{Patk\'o\ifmmode~\check{s}\else
  \v{s}\fi{}}}, \bibinfo{author}{\bibfnamefont{V.~A.} \bibnamefont{Yerokhin}},
  \bibnamefont{and} \bibinfo{author}{\bibfnamefont{K.}~\bibnamefont{Pachucki}},
  \bibinfo{journal}{Phys. Rev. A} \textbf{\bibinfo{volume}{94}},
  \bibinfo{pages}{052508} (\bibinfo{year}{2016}).

\bibitem[{\citenamefont{Patk\'o\ifmmode~\check{s}\else \v{s}\fi{}
  et~al.}(2017)\citenamefont{Patk\'o\ifmmode~\check{s}\else \v{s}\fi{},
  Yerokhin, and Pachucki}}]{patkos17a}
\bibinfo{author}{\bibfnamefont{V.}~\bibnamefont{Patk\'o\ifmmode~\check{s}\else
  \v{s}\fi{}}}, \bibinfo{author}{\bibfnamefont{V.~A.} \bibnamefont{Yerokhin}},
  \bibnamefont{and} \bibinfo{author}{\bibfnamefont{K.}~\bibnamefont{Pachucki}},
  \bibinfo{journal}{Phys. Rev. A} \textbf{\bibinfo{volume}{95}},
  \bibinfo{pages}{012508} (\bibinfo{year}{2017}).

\bibitem[{\citenamefont{Zheng et~al.}(2017)\citenamefont{Zheng, Sun, Chen,
  Jiang, Pachucki, and Hu}}]{zheng17a}
\bibinfo{author}{\bibfnamefont{X.}~\bibnamefont{Zheng}},
  \bibinfo{author}{\bibfnamefont{Y.~R.} \bibnamefont{Sun}},
  \bibinfo{author}{\bibfnamefont{J.-J.} \bibnamefont{Chen}},
  \bibinfo{author}{\bibfnamefont{W.}~\bibnamefont{Jiang}},
  \bibinfo{author}{\bibfnamefont{K.}~\bibnamefont{Pachucki}}, \bibnamefont{and}
  \bibinfo{author}{\bibfnamefont{S.-M.} \bibnamefont{Hu}},
  \bibinfo{journal}{Phys. Rev. Lett.} \textbf{\bibinfo{volume}{119}},
  \bibinfo{pages}{263002} (\bibinfo{year}{2017}).

\bibitem[{\citenamefont{Pastor et~al.}(2006)\citenamefont{Pastor, Giusfredi,
  De~Natale, Hagel, de~Mauro, and Inguscio}}]{pastor06a}
\bibinfo{author}{\bibfnamefont{P.~C.} \bibnamefont{Pastor}},
  \bibinfo{author}{\bibfnamefont{G.}~\bibnamefont{Giusfredi}},
  \bibinfo{author}{\bibfnamefont{P.}~\bibnamefont{De~Natale}},
  \bibinfo{author}{\bibfnamefont{G.}~\bibnamefont{Hagel}},
  \bibinfo{author}{\bibfnamefont{C.}~\bibnamefont{de~Mauro}}, \bibnamefont{and}
  \bibinfo{author}{\bibfnamefont{M.}~\bibnamefont{Inguscio}},
  \bibinfo{journal}{Phys. Rev. Lett.} \textbf{\bibinfo{volume}{97}},
  \bibinfo{pages}{139903} (\bibinfo{year}{2006}).

\bibitem[{\citenamefont{Kim et~al.}(2017)\citenamefont{Kim, Heo, Lee, Park,
  Hong, Hwang, and Yu}}]{kim17a}
\bibinfo{author}{\bibfnamefont{H.}~\bibnamefont{Kim}},
  \bibinfo{author}{\bibfnamefont{M.~S.} \bibnamefont{Heo}},
  \bibinfo{author}{\bibfnamefont{W.~K.} \bibnamefont{Lee}},
  \bibinfo{author}{\bibfnamefont{C.~Y.} \bibnamefont{Park}},
  \bibinfo{author}{\bibfnamefont{H.~G.} \bibnamefont{Hong}},
  \bibinfo{author}{\bibfnamefont{S.~W.} \bibnamefont{Hwang}}, \bibnamefont{and}
  \bibinfo{author}{\bibfnamefont{D.~H.} \bibnamefont{Yu}},
  \bibinfo{journal}{Jpn. J. Appl. Phys.} \textbf{\bibinfo{volume}{56}},
  \bibinfo{pages}{050302} (\bibinfo{year}{2017}).

\bibitem[{\citenamefont{Campbell et~al.}(2017)\citenamefont{Campbell, Hutson,
  Marti, Goban, Darkwah~Oppong, McNally, Sonderhouse, Robinson, Zhang, Bloom
  et~al.}}]{campbell17a}
\bibinfo{author}{\bibfnamefont{S.~L.} \bibnamefont{Campbell}},
  \bibinfo{author}{\bibfnamefont{R.~B.} \bibnamefont{Hutson}},
  \bibinfo{author}{\bibfnamefont{G.~E.} \bibnamefont{Marti}},
  \bibinfo{author}{\bibfnamefont{A.}~\bibnamefont{Goban}},
  \bibinfo{author}{\bibfnamefont{N.}~\bibnamefont{Darkwah~Oppong}},
  \bibinfo{author}{\bibfnamefont{R.~L.} \bibnamefont{McNally}},
  \bibinfo{author}{\bibfnamefont{L.}~\bibnamefont{Sonderhouse}},
  \bibinfo{author}{\bibfnamefont{J.~M.} \bibnamefont{Robinson}},
  \bibinfo{author}{\bibfnamefont{W.}~\bibnamefont{Zhang}},
  \bibinfo{author}{\bibfnamefont{B.~J.} \bibnamefont{Bloom}},
  \bibnamefont{et~al.}, \bibinfo{journal}{Science}
  \textbf{\bibinfo{volume}{358}}, \bibinfo{pages}{90} (\bibinfo{year}{2017}).

\bibitem[{\citenamefont{Notermans et~al.}(2014)\citenamefont{Notermans,
  Rengelink, van Leeuwen, and Vassen}}]{notermans14b}
\bibinfo{author}{\bibfnamefont{R.~P. M. J.~W.} \bibnamefont{Notermans}},
  \bibinfo{author}{\bibfnamefont{R.~J.} \bibnamefont{Rengelink}},
  \bibinfo{author}{\bibfnamefont{K.~A.~H.} \bibnamefont{van Leeuwen}},
  \bibnamefont{and} \bibinfo{author}{\bibfnamefont{W.}~\bibnamefont{Vassen}},
  \bibinfo{journal}{Phys. Rev. A} \textbf{\bibinfo{volume}{90}},
  \bibinfo{pages}{052508} (\bibinfo{year}{2014}).

\bibitem[{\citenamefont{Zhang et~al.}(2016)\citenamefont{Zhang, Tang, Zhang,
  and Shi}}]{zhang16a}
\bibinfo{author}{\bibfnamefont{Y.-H.} \bibnamefont{Zhang}},
  \bibinfo{author}{\bibfnamefont{L.-Y.} \bibnamefont{Tang}},
  \bibinfo{author}{\bibfnamefont{X.-Z.} \bibnamefont{Zhang}}, \bibnamefont{and}
  \bibinfo{author}{\bibfnamefont{T.-Y.} \bibnamefont{Shi}},
  \bibinfo{journal}{Phys. Rev. A} \textbf{\bibinfo{volume}{93}},
  \bibinfo{pages}{052516} (\bibinfo{year}{2016}).

\bibitem[{\citenamefont{Yang et~al.}(2017)\citenamefont{Yang, Mei, Shi, and
  Qiao}}]{yang17a}
\bibinfo{author}{\bibfnamefont{S.-J.} \bibnamefont{Yang}},
  \bibinfo{author}{\bibfnamefont{X.-S.} \bibnamefont{Mei}},
  \bibinfo{author}{\bibfnamefont{T.-Y.} \bibnamefont{Shi}}, \bibnamefont{and}
  \bibinfo{author}{\bibfnamefont{H.-X.} \bibnamefont{Qiao}},
  \bibinfo{journal}{Phys. Rev. A} \textbf{\bibinfo{volume}{95}},
  \bibinfo{pages}{062505} (\bibinfo{year}{2017}).

\bibitem[{\citenamefont{Mohr et~al.}(2012)\citenamefont{Mohr, Taylor, and
  Newell}}]{mohr12a}
\bibinfo{author}{\bibfnamefont{P.~J.} \bibnamefont{Mohr}},
  \bibinfo{author}{\bibfnamefont{B.~N.} \bibnamefont{Taylor}},
  \bibnamefont{and} \bibinfo{author}{\bibfnamefont{D.~B.}
  \bibnamefont{Newell}}, \bibinfo{journal}{Rev. Mod. Phys.}
  \textbf{\bibinfo{volume}{84}}, \bibinfo{pages}{1527} (\bibinfo{year}{2012}).

\bibitem[{\citenamefont{Johnson et~al.}(1988)\citenamefont{Johnson, Blundell,
  and Sapirstein}}]{johnson88a}
\bibinfo{author}{\bibfnamefont{W.~R.} \bibnamefont{Johnson}},
  \bibinfo{author}{\bibfnamefont{S.~A.} \bibnamefont{Blundell}},
  \bibnamefont{and}
  \bibinfo{author}{\bibfnamefont{J.}~\bibnamefont{Sapirstein}},
  \bibinfo{journal}{Phys.~Rev.~A} \textbf{\bibinfo{volume}{37}},
  \bibinfo{pages}{307} (\bibinfo{year}{1988}).

\bibitem[{\citenamefont{{Beloy} and {Derevianko}}(2008)}]{beloy08a}
\bibinfo{author}{\bibfnamefont{K.}~\bibnamefont{{Beloy}}} \bibnamefont{and}
  \bibinfo{author}{\bibfnamefont{A.}~\bibnamefont{{Derevianko}}},
  \bibinfo{journal}{Comp.~Phys.~Commun.} \textbf{\bibinfo{volume}{179}},
  \bibinfo{pages}{310} (\bibinfo{year}{2008}).

\bibitem[{\citenamefont{Pachucki and Sapirstein}(2000)}]{pachucki00a}
\bibinfo{author}{\bibfnamefont{K.}~\bibnamefont{Pachucki}} \bibnamefont{and}
  \bibinfo{author}{\bibfnamefont{J.}~\bibnamefont{Sapirstein}},
  \bibinfo{journal}{Phys. Rev. A} \textbf{\bibinfo{volume}{63}},
  \bibinfo{pages}{012504} (\bibinfo{year}{2000}).

\bibitem[{\citenamefont{Zhang et~al.}(2015)\citenamefont{Zhang, Tang, Zhang,
  and Shi}}]{zhang15}
\bibinfo{author}{\bibfnamefont{Y.-H.} \bibnamefont{Zhang}},
  \bibinfo{author}{\bibfnamefont{L.-Y.} \bibnamefont{Tang}},
  \bibinfo{author}{\bibfnamefont{X.-Z.} \bibnamefont{Zhang}}, \bibnamefont{and}
  \bibinfo{author}{\bibfnamefont{T.-Y.} \bibnamefont{Shi}},
  \bibinfo{journal}{Phys. Rev. A} \textbf{\bibinfo{volume}{92}},
  \bibinfo{pages}{012515} (\bibinfo{year}{2015}).

\bibitem[{\citenamefont{{Yerokhin} and {Pachucki}}(2010)}]{yerokhin10a}
\bibinfo{author}{\bibfnamefont{V.~A.} \bibnamefont{{Yerokhin}}}
  \bibnamefont{and}
  \bibinfo{author}{\bibfnamefont{K.}~\bibnamefont{{Pachucki}}},
  \bibinfo{journal}{\pra} \textbf{\bibinfo{volume}{81}},
  \bibinfo{pages}{022507} (\bibinfo{year}{2010}).

\bibitem[{\citenamefont{Drake and Goldman}(1999)}]{drake99b}
\bibinfo{author}{\bibfnamefont{G.~W.~F.} \bibnamefont{Drake}} \bibnamefont{and}
  \bibinfo{author}{\bibfnamefont{S.~P.} \bibnamefont{Goldman}},
  \bibinfo{journal}{Can. J. Phys.} \textbf{\bibinfo{volume}{77}},
  \bibinfo{pages}{835} (\bibinfo{year}{1999}).

\bibitem[{\citenamefont{Drake}(2006)}]{drake06a}
\bibinfo{author}{\bibfnamefont{G.~W.~F.} \bibnamefont{Drake}},
  \emph{\bibinfo{title}{Handbook of atomic, molecular, and optical physics}}
  (\bibinfo{publisher}{Springer, New York}, \bibinfo{year}{2006}).

\bibitem[{\citenamefont{{Drake} and {Morton}}(2007)}]{drake07a}
\bibinfo{author}{\bibfnamefont{G.~W.~F.} \bibnamefont{{Drake}}}
  \bibnamefont{and} \bibinfo{author}{\bibfnamefont{D.~C.}
  \bibnamefont{{Morton}}}, \bibinfo{journal}{Astrophys. J. Suppl. Ser.}
  \textbf{\bibinfo{volume}{170}}, \bibinfo{pages}{251} (\bibinfo{year}{2007}).

\bibitem[{\citenamefont{for~other energies and matrix elements of~helium
  isotopes}(2018)}]{supp18a}
\bibinfo{author}{\bibfnamefont{S.~S.~M.} \bibnamefont{for~other energies}}
  \bibnamefont{and} \bibinfo{author}{\bibnamefont{matrix elements of~helium
  isotopes}} (\bibinfo{year}{2018}).

\bibitem[{\citenamefont{Morton et~al.}(2006{\natexlab{b}})\citenamefont{Morton,
  Wu, and Drake}}]{morton06b}
\bibinfo{author}{\bibfnamefont{D.~C.} \bibnamefont{Morton}},
  \bibinfo{author}{\bibfnamefont{Q.}~\bibnamefont{Wu}}, \bibnamefont{and}
  \bibinfo{author}{\bibfnamefont{G.~W.~F.} \bibnamefont{Drake}},
  \bibinfo{journal}{Can. J. Phys.} \textbf{\bibinfo{volume}{84}},
  \bibinfo{pages}{83} (\bibinfo{year}{2006}{\natexlab{b}}).

\bibitem[{\citenamefont{Jiang and Mitroy}(2013)}]{jiang13c}
\bibinfo{author}{\bibfnamefont{J.}~\bibnamefont{Jiang}} \bibnamefont{and}
  \bibinfo{author}{\bibfnamefont{J.}~\bibnamefont{Mitroy}},
  \bibinfo{journal}{Phys. Rev. A} \textbf{\bibinfo{volume}{88}},
  \bibinfo{pages}{032505} (\bibinfo{year}{2013}).

\bibitem[{\citenamefont{{Rengelink} et~al.}(2018)\citenamefont{{Rengelink},
  {van der Werf}, {Notermans}, {Jannin}, {Eikema}, {Hoogerland}, and
  {Vassen}}}]{wim18a}
\bibinfo{author}{\bibfnamefont{R.~J.} \bibnamefont{{Rengelink}}},
  \bibinfo{author}{\bibfnamefont{Y.}~\bibnamefont{{van der Werf}}},
  \bibinfo{author}{\bibfnamefont{R.~P.~M.~J.~W.} \bibnamefont{{Notermans}}},
  \bibinfo{author}{\bibfnamefont{R.}~\bibnamefont{{Jannin}}},
  \bibinfo{author}{\bibfnamefont{K.~S.~E.} \bibnamefont{{Eikema}}},
  \bibinfo{author}{\bibfnamefont{M.~D.} \bibnamefont{{Hoogerland}}},
  \bibnamefont{and} \bibinfo{author}{\bibfnamefont{W.}~\bibnamefont{{Vassen}}},
  \bibinfo{journal}{ArXiv e-prints}  (\bibinfo{year}{2018}),
  \eprint{1804.06693}.

\bibitem[{\citenamefont{Rengelink et~al.}(2018)\citenamefont{Rengelink, van~der
  Werf, Notermans, Jannin, Eikema, Hoogerland, and Vassen}}]{wim18b}
\bibinfo{author}{\bibfnamefont{R.~J.} \bibnamefont{Rengelink}},
  \bibinfo{author}{\bibfnamefont{Y.}~\bibnamefont{van~der Werf}},
  \bibinfo{author}{\bibfnamefont{R.~P. M. J.~W.} \bibnamefont{Notermans}},
  \bibinfo{author}{\bibfnamefont{R.}~\bibnamefont{Jannin}},
  \bibinfo{author}{\bibfnamefont{K.~S.~E.} \bibnamefont{Eikema}},
  \bibinfo{author}{\bibfnamefont{M.~D.} \bibnamefont{Hoogerland}},
  \bibnamefont{and} \bibinfo{author}{\bibfnamefont{W.}~\bibnamefont{Vassen}},
  \bibinfo{journal}{Nature Physics} (\bibinfo{year}{2018}).

\end{thebibliography}

\end{document}


\title{ Supplemental Material for

Relativistic full-configuration-interaction calculations of magic wavelengths for the $2\,^3S_1\rightarrow2\,^1S_0$ transition of helium isotopes }

\author{Fang-Fei Wu$^{1,2}$, San-Jiang Yang$^{1,3}$, Yong-Hui Zhang$^{1,*}$~\footnotetext{*Email Address: yhzang@wipm.ac.cn}, Jun-Yi Zhang$^{1}$, Hao-Xue Qiao$^3$, Ting-Yun Shi$^{1,4}$, and Li-Yan Tang$^{1,\dag}$~\footnotetext{\dag Email Address: lytang@wipm.ac.cn}}

\affiliation {$^1$State Key Laboratory of Magnetic Resonance and
Atomic and Molecular Physics, Wuhan Institute of Physics and
Mathematics, Chinese Academy of Sciences, Wuhan 430071, People's Republic of China}

\affiliation {$^2$University of Chinese Academy of Sciences, Beijing 100049, People's Republic of China}

\affiliation {$^3$Department of Physics, Wuhan University, Wuhan 430072, People's Republic of China}

\affiliation {$^4$ Center for Cold Atom Physics, Chinese Academy of Sciences,
Wuhan 430071, People¡¯s Republic of China}

\date{\today}

\pacs{31.15.ap, 31.15.ac, 32.10.Dk} \maketitle

Table~\ref{t1} -~\ref{t4} present the comparison of the energies of helium between present RCI calculations and Hylleraas calculations~\cite{drake06a}. Table~\ref{t7} -~\ref{t8} present the comparison of the reduced matrix elements between present RCI calculations and Hylleraas calculations~\cite{drake06a}. All the values in these tables are obtained from one diagonalization of Hamiltonian. These results are presented as a benchmark for comparison with experiment and theory.

\begin{table*}[!htbp]
\caption{\label{t1} Comparison of the energies (in a.u.) for $n\,^1S_0\,(n \leq 13)$ states of $^4$He and $^3$He. The numbers in parentheses are computational uncertainties. }
\begin{ruledtabular}
\begin{tabular}{clcll}
 \multicolumn{1}{c}{State}
&\multicolumn{2}{c}{$^4$He} &\multicolumn{2}{c}{$^3$He}\\
\cline{2-3} \cline{4-5}
&\multicolumn{1}{c}{RCI} &\multicolumn{1}{c}{Hylleraas~\cite{drake06a}}
&\multicolumn{1}{c}{RCI} &\multicolumn{1}{c}{Hylleraas~\cite{drake06a}}\\
\hline
$1\,^1S_0$  	&	$-$2.903 35(2)     	&	$-$2.903 408 72 	&	$-$2.903 22(2)	    &	$-$2.903 271 37	\\
$2\,^1S_0$  	&	$-$2.145 785(1) 	&	$-$2.145 787 13 	&	$-$2.145 687(1)	&	$-$2.145 690 46	\\
$3\,^1S_0$  	&	$-$2.061 095(1)  	&	$-$2.061 096 44 	&	$-$2.061 003(1)	&	$-$2.061 003 88	\\
$4\,^1S_0$  	&	$-$2.033 414 1(2) 	&	$-$2.033 414 83 	&	$-$2.033 322 8(2)	&	$-$2.033 323 58	\\
$5\,^1S_0$  	&	$-$2.021 006 1(2) 	&	$-$2.021 006 60 	&	$-$2.020 915 4(2)	&	$-$2.020 915 93	\\
$6\,^1S_0$  	&	$-$2.014 393 3(2) 	&	$-$2.014 393 72 	&	$-$2.014 302 9(2)	&	$-$2.014 303 36	\\
$7\,^1S_0$  	&	$-$2.010 456 5(2) 	&	$-$2.010 456 92 	&	$-$2.010 366 3(2)	&	$-$2.010 366 75	\\
$8\,^1S_0$  	&	$-$2.007 924 7(2) 	&	$-$2.007 925 10 	&	$-$2.007 834 6(2)	&	$-$2.007 835 05	\\
$9\,^1S_0$  	&	$-$2.006 200 9(2) 	&	$-$2.006 201 27 	&	$-$2.006 110 9(2)	&	$-$2.006 111 29	\\
$10\,^1S_0$ 	&	$-$2.004 974 5(2) 	&	$-$2.004 974 87 	&	$-$2.004 884 6(2)	&	$-$2.004 884 94	\\
$11\,^1S_0$ 	&	$-$2.004 071 1(2) 	&		                &	$-$2.003 981 2(2)	&		\\
$12\,^1S_0$ 	&	$-$2.003 386 4(2) 	&		                &	$-$2.003 296 5(2)	&		\\
$13\,^1S_0$ 	&	$-$2.002 855 1(2) 	&		                &	$-$2.002 765 3(2)	&		\\
\end{tabular}
\begin{tablenotes}
\item[] $^a$ Both values are obtained by increasing the partial wave of $l_{max}$ to 20.
\end{tablenotes}
\end{ruledtabular}
\end{table*}

\begin{table*}[!htbp]
\caption{\label{t3} The same as Table~\ref{t1}, but for $n\,^3S_1\,(n \leq 13)$ states of $^4$He and $^3$He. }
\begin{ruledtabular}
\begin{tabular}{clcll}
 \multicolumn{1}{c}{State}
&\multicolumn{2}{c}{$^4$He} &\multicolumn{2}{c}{$^3$He}\\
\cline{2-3} \cline{4-5}
&\multicolumn{1}{c}{RCI} &\multicolumn{1}{c}{Hylleraas~\cite{drake06a}}
&\multicolumn{1}{c}{RCI} &\multicolumn{1}{c}{Hylleraas~\cite{drake06a}}\\
\hline
$2\,^3S_1$  	&	$-$2.175 045 25(2) 	&	$-$2.175 045 67 	&	$-$2.174 947 30(2)	&	$-$2.174 947 79	\\
$3\,^3S_1$  	&	$-$2.068 514 02(2) 	&	$-$2.068 514 37 	&	$-$2.068 421 12(2)	&	$-$2.068 421 51	\\
$4\,^3S_1$  	&	$-$2.036 340 19(2) 	&	$-$2.036 340 52 	&	$-$2.036 248 79(2)	&	$-$2.036 249 16	\\
$5\,^3S_1$  	&	$-$2.022 448 45(2) 	&	$-$2.022 448 78 	&	$-$2.022 357 69(2)	&	$-$2.022 358 06	\\
$6\,^3S_1$  	&	$-$2.015 207 84(2) 	&	$-$2.015 208 17 	&	$-$2.015 117 41(2)	&	$-$2.015 117 78	\\
$7\,^3S_1$  	&	$-$2.010 960 80(2) 	&	$-$2.010 961 12 	&	$-$2.010 870 56(2)	&	$-$2.010 870 92	\\
$8\,^3S_1$  	&	$-$2.008 258 32(2) 	&	$-$2.008 258 64 	&	$-$2.008 168 21(2)	&	$-$2.008 168 57	\\
$9\,^3S_1$  	&	$-$2.006 432 93(2) 	&	$-$2.006 433 26 	&	$-$2.006 342 90(2)	&	$-$2.006 343 26	\\
$10\,^3S_1$ 	&	$-$2.005 142 37(2) 	&	$-$2.005 142 69 	&	$-$2.005 052 40(2)	&	$-$2.005 052 76	\\
$11\,^3S_1$ 	&	$-$2.004 196 39(2) 	&		                &	$-$2.004 106 46(2)	&		\\
$12\,^3S_1$ 	&	$-$2.003 482 41(4)  &		                &	$-$2.003 392 51(4)	&		\\
$13\,^3S_1$ 	&	$-$2.002 930 32(4)  &		                &	$-$2.002 840 45(4)	    &		\\
\end{tabular}
\end{ruledtabular}
\end{table*}

\begin{table*}[!htbp]
\caption{\label{t4} The same as Table~\ref{t1}, for $n\,^3P_J\,(n \leq 13)$ states of $^4$He and $^3$He. The numbers in parentheses are computational uncertainties.}
\begin{ruledtabular}
\begin{tabular}{clcll}
 \multicolumn{1}{c}{State}
&\multicolumn{2}{c}{$^4$He} &\multicolumn{2}{c}{$^3$He}\\
\cline{2-3} \cline{4-5}
&\multicolumn{1}{c}{RCI} &\multicolumn{1}{c}{Hylleraas~\cite{drake06a}}
&\multicolumn{1}{c}{RCI} &\multicolumn{1}{c}{Hylleraas~\cite{drake06a}}\\
\hline
$2\,^3P_0$  	&	$-$2.132 980 94(2)	&	$-$2.132 981 43 	&	$-$2.132 888 13(2)	&	$-$2.132 888 66	\\
$3\,^3P_0$  	&	$-$2.057 906 35(2)	&	$-$2.057 906 72 	&	$-$2.057 814 84(2)	&	$-$2.057 815 25 \\
$4\,^3P_0$  	&	$-$2.032 152 61(2)	&	$-$2.032 152 95 	&	$-$2.032 061 77(2)	&	$-$2.032 062 15	\\
$5\,^3P_0$  	&	$-$2.020 380 85(2)	&	$-$2.020 381 18 	&	$-$2.020 290 37(2)	&	$-$2.020 290 74	\\
$6\,^3P_0$  	&	$-$2.014 038 40(2)	&	$-$2.014 038 73 	&	$-$2.013 948 13(2)	&	$-$2.013 948 41	\\
$7\,^3P_0$  	&	$-$2.010 235 87(2)	&	$-$2.010 236 20 	&	$-$2.010 145 73(2)	&	$-$2.010 146 10	\\
$8\,^3P_0$  	&	$-$2.007 778 23(2)	&	$-$2.007 778 56 	&	$-$2.007 688 19(2)	&	$-$2.007 688 55	\\
$9\,^3P_0$  	&	$-$2.006 098 70(2)	&	$-$2.006 099 02 	&	$-$2.006 008 72(2)	&	$-$2.006 009 08	\\
$10\,^3P_0$ 	&	$-$2.004 900 39(2)	&	$-$2.004 900 71 	&	$-$2.004 810 45(2)	&	$-$2.004 810 82	\\
$11\,^3P_0$ 	&	$-$2.004 015 60(2)	&		                &	$-$2.003 925 70(2)	&		\\
$12\,^3P_0$ 	&	$-$2.003 343 79(4) 	&		                &	$-$2.003 253 92(4)	&		\\
$13\,^3P_0$ 	&	$-$2.002 821 71(4) 	&		                &	$-$2.002 731 86(4)	&		\\
\hline
$2\,^3P_1$  	&	$-$2.132 985 44(2)	&	$-$2.132 985 93 	&	$-$2.132 892 63(2)	&	$-$2.132 893 16	\\
$3\,^3P_1$  	&	$-$2.057 907 58(2)	&	$-$2.057 907 95 	&	$-$2.057 816 07(2)	&	$-$2.057 816 48	\\
$4\,^3P_1$  	&	$-$2.032 153 12(2)	&	$-$2.032 153 46 	&	$-$2.032 062 27(2)	&	$-$2.032 062 65	\\
$5\,^3P_1$  	&	$-$2.020 381 10(2)	&	$-$2.020 381 44 	&	$-$2.020 290 62(2)	&	$-$2.020 290 99	\\
$6\,^3P_1$  	&	$-$2.014 038 54(2)	&	$-$2.014 038 87 	&	$-$2.013 948 27(2)	&	$-$2.013 948 55	\\
$7\,^3P_1$  	&	$-$2.010 235 96(2)	&	$-$2.010 236 29 	&	$-$2.010 145 82(2)	&	$-$2.010 146 19	\\
$8\,^3P_1$  	&	$-$2.007 778 29(2)	&	$-$2.007 778 62 	&	$-$2.007 688 25(2)	&	$-$2.007 688 61 \\
$9\,^3P_1$  	&	$-$2.006 098 74(2)	&	$-$2.006 099 07 	&	$-$2.006 008 76(2)	&	$-$2.006 009 12	\\
$10\,^3P_1$ 	&	$-$2.004 900 42(2)	&	$-$2.004 900 75 	&	$-$2.004 810 48(2)	&	$-$2.004 810 85	\\
$11\,^3P_1$ 	&	$-$2.004 015 63(2)	&		                &	$-$2.003 925 72(2)	&		\\
$12\,^3P_1$ 	&	$-$2.003 343 81(4) 	&		                &	$-$2.003 253 93(4)	&		\\
$13\,^3P_1$ 	&	$-$2.002 821 73(4) 	&		                &	$-$2.002 731 87(4)	&		\\
\hline
$2\,^3P_2$  	&	$-$2.132 985 74(2) 	&	$-$2.132 986 27 	&	$-$2.132 892 97(2)	&	$-$2.132 893 51	\\
$3\,^3P_2$  	&	$-$2.057 907 67(2) 	&	$-$2.057 908 06 	&	$-$2.057 816 17(2)	&	$-$2.057 816 58	\\
$4\,^3P_2$  	&	$-$2.032 153 15(2) 	&	$-$2.032 153 50 	&	$-$2.032 062 32(2)	&	$-$2.032 062 69	\\
$5\,^3P_2$  	&	$-$2.020 381 12(2) 	&	$-$2.020 381 46 	&	$-$2.020 290 64(2)	&	$-$2.020 291 01	\\
$6\,^3P_2$  	&	$-$2.014 038 55(2) 	&	$-$2.014 038 88 	&	$-$2.013 948 28(2)	&	$-$2.013 948 57	\\
$7\,^3P_2$  	&	$-$2.010 235 97(2) 	&	$-$2.010 236 29 	&	$-$2.010 145 84(2)	&	$-$2.010 146 20	\\
$8\,^3P_2$  	&	$-$2.007 778 30(2) 	&	$-$2.007 778 62 	&	$-$2.007 688 25(2)	&	$-$2.007 688 61	\\
$9\,^3P_2$  	&	$-$2.006 098 74(2) 	&	$-$2.006 099 07 	&	$-$2.006 008 76(2)	&	$-$2.006 009 12	\\
$10\,^3P_2$ 	&	$-$2.004 900 42(2) 	&	$-$2.004 900 75 	&	$-$2.004 810 49(2)	&	$-$2.004 810 85	\\
$11\,^3P_2$ 	&	$-$2.004 015 63(2) 	&		                &	$-$2.003 925 72(2)	&		\\
$12\,^3P_2$ 	&	$-$2.003 343 81(4) 	&		                &	$-$2.003 253 94(4)	&		\\
$13\,^3P_2$ 	&	$-$2.002 821 73(4)  &		                &	$-$2.002 731 79(4)  &		\\
\end{tabular}
\end{ruledtabular}
\end{table*}

\begin{table*}[!htbp]
\caption{\label{t7} Comparison of reduced matrix elements between present RCI calculations and Hylleraas calculations~\cite{drake07a} of the $2\,^3S_1\rightarrow n\,^3P_J (n \leq 13)$ transitions for helium. The numbers in parentheses are computational uncertainties.}
\begin{ruledtabular}
\begin{tabular}{clllclll}
$n\,^3P_J$ &\multicolumn{2}{c}{RCI} &\multicolumn{1}{c}{Hylleraas~\cite{drake07a}} & $n\,^3P_J$ &\multicolumn{2}{c}{RCI}  &\multicolumn{1}{c}{Hylleraas~\cite{drake07a}}\\
\cline{2-3} \cline{4-4}  \cline{6-7}  \cline{8-8}
& \multicolumn{1}{c}{$^4He$} &\multicolumn{1}{c}{$^3He$}  &  \multicolumn{1}{c}{$^4He$}   & & \multicolumn{1}{c}{$^4He$} &\multicolumn{1}{c}{$^3He$}  &  \multicolumn{1}{c}{$^4He$}   \\
\hline
$2\,^3P_2$  &5.660 70(2) &5.660 906(3) &5.660 605  &$8\,^3P_2$  &0.205 91(2) &0.205 938(2) &0.205 885  \\
$2\,^3P_1$  &4.384 77(2) &4.384 920(2) &4.384 692  &$8\,^3P_1$  &0.159 50(2) &0.159 517(2) &0.159 476  \\
$2\,^3P_0$  &2.531 58(2) &2.531 673(2) &2.531 375  &$8\,^3P_0$  &0.092 08(2) &0.092 087(2) &0.092 073  \\
$3\,^3P_2$  &1.173 17(2) &1.173 398(2) &1.172 934  &$9\,^3P_2$  &0.170 76(2) &0.170 785(2) &0.170 742  \\
$3\,^3P_1$  &0.908 71(2) &0.908 893(2) &0.908 546  &$9\,^3P_1$  &0.132 27(2) &0.132 288(2) &0.132 254  \\
$3\,^3P_0$  &0.524 54(2) &0.524 644(2) &0.524 549  &$9\,^3P_0$  &0.076 36(2) &0.076 368(2) &0.076 356  \\
$4\,^3P_2$  &0.671 56(2) &0.671 667(2) &0.671 450  &$10\,^3P_2$ &0.144 70(2) &0.144 713(2) &0.144 674  \\
$4\,^3P_1$  &0.520 18(2) &0.520 264(2) &0.520 103  &$10\,^3P_2$ &0.112 08(2) &0.112 093(2) &0.112 064  \\
$4\,^3P_0$  &0.300 34(2) &0.300 331(2) &0.300 281  &$10\,^3P_0$ &0.064 70(2) &0.064 710(2) &0.064 701  \\
$5\,^3P_2$  &0.449 40(2) &0.449 465(2) &0.449 332  &$11\,^3P_2$ &0.124 71(2) &0.124 720(2) & \\
$5\,^3P_1$  &0.348 10(2) &0.348 149(2) &0.348 050  &$11\,^3P_1$ &0.096 59(2) &0.096 607(2) &\\
$5\,^3P_0$  &0.200 99(2) &0.200 979(2) &0.200 944  &$11\,^3P_0$ &0.055 76(2) &0.055 770(2)  & \\
$6\,^3P_2$  &0.329 31(2) &0.329 356(2) &0.329 269  &$12\,^3P_2$ &0.108 97(2) &0.108 978(2)  & \\
$6\,^3P_1$  &0.255 08(2) &0.255 115(2) &0.255 047  &$12\,^3P_1$ &0.084 40(2) &0.084 413(2)   & \\
$6\,^3P_0$  &0.147 28(2) &0.147 273(2) &0.147 252  &$12\,^3P_0$ &0.048 72(2) &0.048 731(2)  & \\
$7\,^3P_2$  &0.255 40(2) &0.255 437(2) &0.255 365  &$13\,^3P_2$ &0.096 30(2) &0.096 313(2)  & \\
$7\,^3P_1$  &0.197 83(2) &0.197 858(2) &0.197 806  &$13\,^3P_1$ &0.074 59(2) &0.074 603(2)  & \\
$7\,^3P_0$  &0.114 23(2) &0.114 220(2) &0.114 201  &$13\,^3P_0$ &0.043 06(2) &0.043 067(3) & \\
\end{tabular}
\end{ruledtabular}
\end{table*}

\begin{table*}[!htbp]
\caption{\label{t8} The same as Table~\ref{t7}, but for the $2\,^1S_0\rightarrow n\,^3P_1$ and $2\,^3S_1\rightarrow n\,^1P_1 (n \leq 10)$ forbidden transitions of helium. The numbers in the square brackets denote powers of ten.}
\begin{ruledtabular}
\begin{tabular}{clllclll}
 \multicolumn{1}{c}{$n\,^3P_1$}
&\multicolumn{2}{c}{RCI}             &\multicolumn{1}{c}{Hylleraas~\cite{morton15}}
&\multicolumn{1}{c}{$ n\,^1P_1$} &\multicolumn{2}{c}{RCI} &\multicolumn{1}{c}{Hylleraas~\cite{morton15}}  \\
\cline{2-3} \cline{4-4}  \cline{6-7}  \cline{8-8}
& \multicolumn{1}{c}{$^4He$} &\multicolumn{1}{c}{$^3He$}  &  \multicolumn{1}{c}{$^4He$}   & & \multicolumn{1}{c}{$^4He$} &\multicolumn{1}{c}{$^3He$}  &  \multicolumn{1}{c}{$^4He$}   \\
\hline
$2\,^3P_1$  &1.34(2)[--3]    &1.34(1)[--3]  &1.337[--3]  &$2\,^1P_1$  &1.266(3)[--3]   &1.268(2)[--3]  &1.262[--3] \\
$3\,^3P_1$  &5.41(4)[--4]    &5.44(1)[--4]  &5.48[--4]   &$3\,^1P_1$  &1.21(2)[--4]    &1.22(2)[--4]   &1.21[--4] \\
$4\,^3P_1$  &2.43(4)[--4]    &2.46(1)[--4]  &            &$4\,^1P_1$  &8.12(3)[--5]    &8.14(2)[--5]   & \\
$5\,^3P_1$  &1.50(2)[--4]    &1.52(1)[--4]  &            &$5\,^1P_1$  &5.72(3)[--5]    &5.70(2)[--5]   & \\
$6\,^3P_1$  &1.06(2)[--4]    &1.07(1)[--4]  &            &$6\,^1P_1$  &4.28(2)[--5]    &4.26(2)[--5]   & \\
$7\,^3P_1$  &8.0(2)[--5]     &8.12(2)[--5]  &            &$7\,^1P_1$  &3.35(2)[--5]    &3.34(2)[--5]   & \\
$8\,^3P_1$  &6.3(2)[--5]     &6.46(2)[--5]  &            &$8\,^1P_1$  &2.71(2)[--5]    &2.72(2)[--5]   & \\
$9\,^3P_1$  &5.2(2)[--5]     &5.31(2)[--5]  &            &$9\,^1P_1$  &2.27(2)[--5]    &2.26(2)[--5]   & \\
$10\,^3P_1$ &4.3(2)[--5]     &4.47(2)[--5]  &            &$10\,^1P_1$ &1.94(3)[--5]    &1.92(2)[--5]   & \\
\end{tabular}
\end{ruledtabular}
\end{table*}

\clearpage